\def\tipo{2}

\ifnum \tipo = 1
 \documentstyle[prl,twocolumn,rotate,aps,psfig]{revtex}
 
 \def \frontmatter{\twocolumn[\hsize\textwidth\columnwidth\hsize\csname@twocolumnfalse\endcsname}

\fi
\ifnum \tipo = 2
 \documentstyle[prl,multicol,rotate,aps,psfig]{revtex}

 \def\frontmatter{}
 
 \fi
\ifnum \tipo = 3
 \documentstyle[prl,manuscript,rotate,aps,psfig]{revtex}
 
 \def\frontmatter{}
 
\fi

\begin{document}
\draft

\input epsf.sty
\frontmatter

\title{Noise Effects on the Complex Patterns of Abnormal Heartbeats}

\author{Verena~Schulte-Frohlinde$^{1}$, 
        Yosef~Ashkenazy$^{1}$,
        Plamen~Ch.~Ivanov$^{1,2}$,\\
        Leon~Glass$^{3}$, 
        Ary~L.~Goldberger$^{2}$, and
        H.~Eugene~Stanley$^1$}

\address{ $^1$ Center for Polymer Studies, Department of Physics,
  		Boston University, Boston, MA 02215\\
          $^2$ Cardiovascular Division, Harvard Medical School,
		Beth Israel Deaconess Medical Center, Boston, MA 02215\\
	  $^3$ Department of Physiology, McGill University, Montreal, Quebec, H3G IY6}

\date{\today}

\maketitle
\begin{abstract}
Patients at high risk for sudden death often exhibit complex heart
rhythms in which abnormal heartbeats are interspersed with normal
heartbeats. We analyze such a complex rhythm in a single patient over
a 12-hour period and show that the rhythm can be described by a
theoretical model consisting of two interacting oscillators with
stochastic elements.  By varying the magnitude of the noise, we show
that for an intermediate level of noise, the model gives best
agreement with key statistical features of the dynamics.

\end{abstract}


\ifnum \tipo = 1
 ]
 \narrowtext
\fi
\ifnum \tipo = 2
 \begin{multicols}{2}
\fi

Individuals with frequent abnormal heartbeats (Fig.~\ref{data}) may be
at high risk for sudden cardiac death~\cite{echt}.  Such abnormal
heart rhythms often have a random appearance.  Attempts have been made
to analyze these rhythms by inspecting short data
strips~\cite{moe,courte} and matching them beat by beat to various
models~\cite{courte,jalife,glass,glassb}.  Other
approaches~\cite{depaola,liebovitch,babloyantz} characterize
statistical properties in longer records of up to several hours.
However, the mechanisms underlying these abnormal rhythms and their
changes over long periods of time remain elusive.  Here we show that a
theoretical model consisting of two coupled
oscillators~\cite{courte,jalife,glass,glassb,depaola} describes the observed
patterns of abnormal heartbeats in one clinical case provided we
introduce noise to the periods of the oscillators.  
This approach may generalize to the analysis of
the underlying mechanism of a large number of records with complex
patterns of abnormal heartbeats.

We consider a continuous 12-hour segment of the ambulatory
electrocardiographic record of an individual with heart failure and
frequent abnormal heartbeats (Fig.~\ref{data}).  The normal sinus
heartbeats, S-beats, arise from activity in the sinus node, the normal
pacemaker of the heart.  The time intervals between the normal beats
appear to be periodic in this short tracing, but they do fluctuate during
the 12-hour period.  The abnormal ventricular beats, V-beats, arise in
the lower chambers of the heart, the ventricles.  Although the timing
between the V-beats appears to be irregular, the histogram of the
interventricular time intervals, the time intervals between
consecutive V-beats, consists of {\it equidistant} peaks implying that
the interventricular time intervals are multiples of a fixed number
(Fig.~\ref{data}).  This result is consistent with the possibility
that the heart rhythm arises from a competition between two
oscillators: the normal sinus oscillator, and an abnormal ventricular
oscillator with periods $T_S$ and $T_V$, respectively.  Such rhythms
are called parasystolic.

After each S-beat or V-beat, there is a time period, 
called the refractory time $\theta$, during which all other heartbeats 
are blocked. 
Thus, all V-beats must occur at least a time $\theta$ after an S-beat.
Because in our case $\theta/T_S \geq 1/2$, such a V-beat
will block the following S-beat\cite{refract}.
\def\figuraI{
\begin{figure}[t!]
\vspace*{.0cm}
\centerline{ 
\vbox{  
    \hbox{
         \epsfxsize=8.2cm  \epsfbox{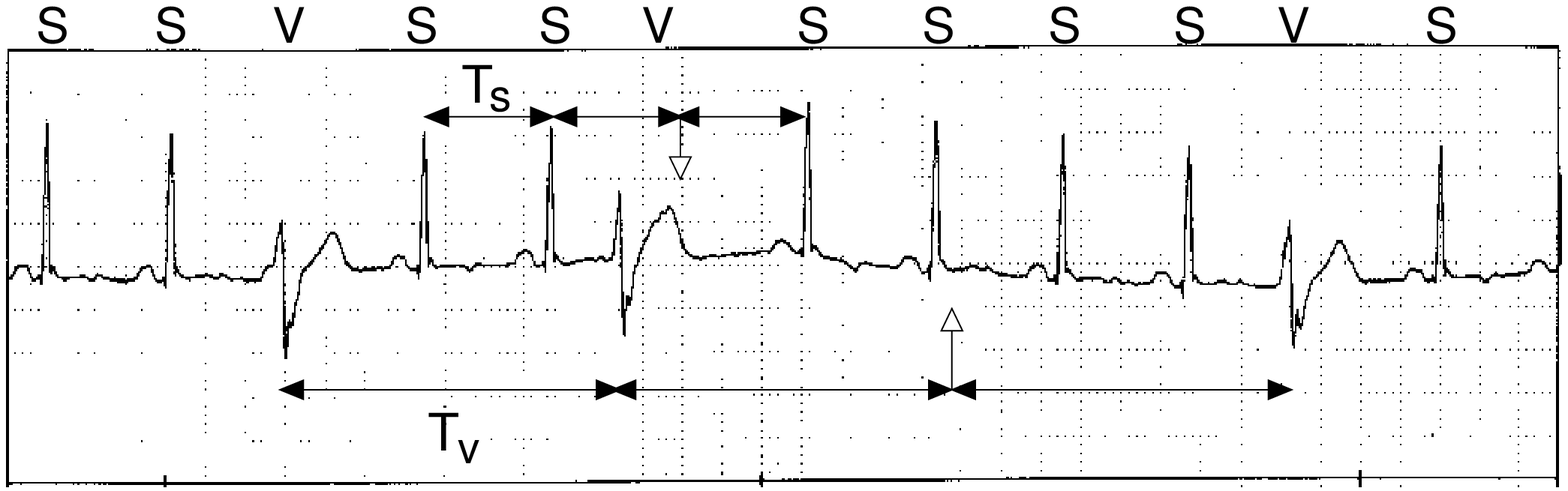}
         } } }
\vspace*{0.cm}
\centerline{
\vbox{
    \hbox{\hspace*{-0.1cm}
         \epsfxsize=8.5cm \epsfbox{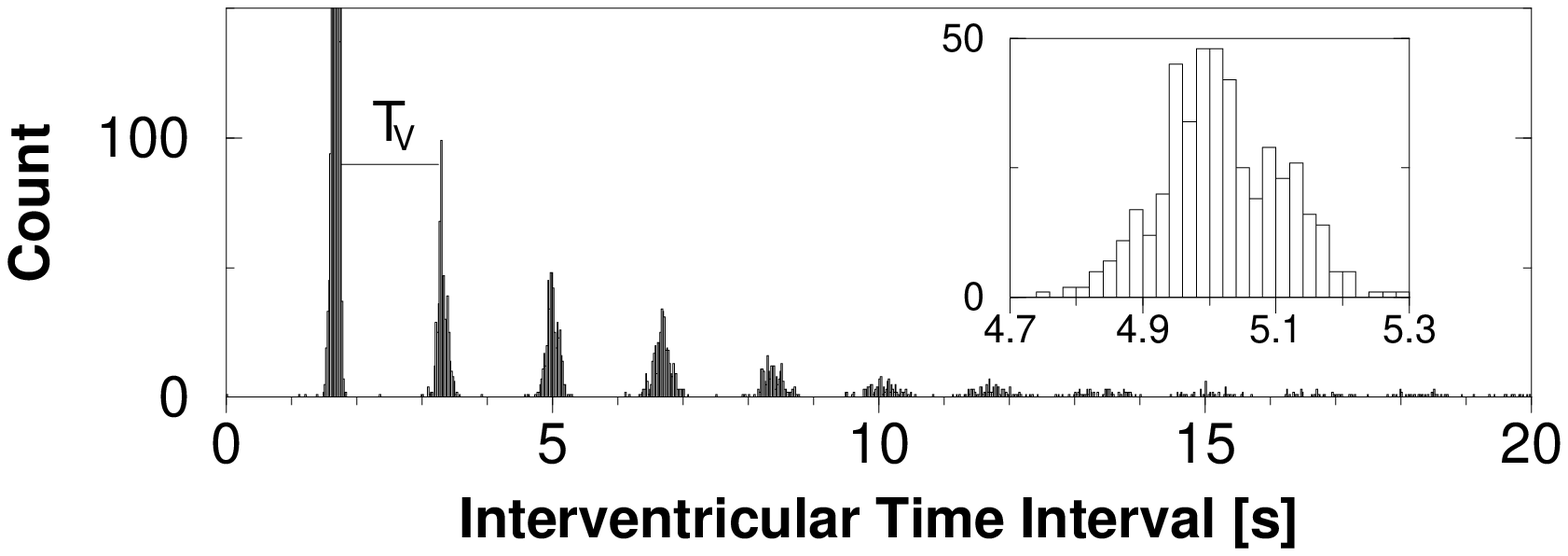}
         } } }
\vspace*{0.0cm}
\caption{Upper panel: Electrocardiogram over 6.6 s of a patient with
heart failure. Sinus (S) and ventricular (V) beats differ in morphology.
S-beats are separated by an interval $T_S$. 
The time interval between the S-beats 
with an intervening V-beat is approximately $2\,T_S$
implying that an S-beat has been blocked (downward white arrow).
The interventricular time interval $T_V$,  determined from the histogram
in the lower panel, is also indicated and is consistent with 
a blocked V-beat (upward white arrow).
Lower panel: 
The histogram of the time intervals between 
consecutive abnormal V-beats for the 12-hour record 
from which the segment in the upper panel was extracted
($\approx 80\,000$ beats with $\approx 5000$ V-beats).
The time interval between the peaks is $1.67$~s ($\approx 3\,T_S$).
The inset is an enlargement of the peak at 5 s.
The sampling rate of the underlying electrocardiographic 
time series is 128~Hz.}
\label{data}
\end{figure}
}
\ifnum\tipo=1
  \figuraI
\fi
\ifnum\tipo=2
  \figuraI
\fi
\noindent
To describe the parasystole mechanism 
we introduce the phase $\phi_i$
of the $i$th ventricular beat in the sinus cycle, i.e., the time interval 
between the $i$th ventricular beat and the previous sinus beat 
normalized by $T_S$\cite{courte,jalife,glass,glassb}.  
Successive values of $\phi_i$ are determined
by iterating the difference equation: 
\begin{equation}
\label{fde}
 \phi_{i+1} = (\phi_{i} + T_V/T_S )~{\rm mod}\,1,
\end{equation}
where a ventricular beat is expressed if $\phi_{i+1}>\theta/T_S$.
For a fixed irrational ratio $T_V/T_S$  (incommensurate periods),
the phase of the V-oscillator with respect to the S-oscillator 
will, for sufficiently long times, be equally distributed in the
interval [0,1].
Interestingly, the sequence of the number of intervening S-beats 
between consecutive V-beats, called the NIB sequence,
consists of only three different values (NIB-triplet).
The values in these NIB-triplets change with 
$T_V/T_S$ and $\theta/T_S$~\cite{rules}.  
Examples are shown in Fig.~\ref{cartoon}a.

The uniform distribution of the phases of the V-beats  allows us
to derive the fraction of V-beats. 
Call $n_V$, the number of V-beats observed on the electrocardiogram 
in a time interval $T$, and $N$ the total number of S-beats and V-beats
during the same time interval.   
The total number of observed and blocked V-beats in $T$ is $T/T_V$ and
the total number of observed beats is $N=T/T_S$. 
The fraction of observed V-beats, i.e. those which  
occur outside of the refractory period
is $(T_S-\theta)/{T_S}$. 
It follows that $n_V=(T_S-\theta)/{T_S}\times T/T_V$, and consequently
\begin{equation}
{n_V}/{N}=(T_S-\theta)/{T_V}.
\label{eq_V_freq}
\end{equation}
Perturbation of the timing of the V-beats by Gaussian noise does not
affect this distribution,
because the uniform distribution of the phases of the V-beats 
remains unchanged.

To illustrate the
properties of the unperturbed model, we iterate Eq.~(\ref{fde}) 
using typical parameter values for the patient under consideration.  
Histograms showing the NIB values for three sets of parameters are
shown in Fig. \ref{nib}a. 

We now compare the results of Eq.~(\ref{fde}) with the clinical data. 
Since the value of $T_S$ fluctuates during the record, we combine the
data from different times of the day during which $T_S$ falls in 
a fixed 10 ms range. Fig.~\ref{nib}b shows the histograms of NIB
values for three different values of $T_S$. 
For $T_S$=0.61~s ($T_V/T_S<3$) and 
$T_S$=0.51~s ($T_V/T_S>3$),
we find the same NIB-triplets
as in the deterministic model (Fig.~\ref{nib}a).
However, additional peaks in the data (Fig.~\ref{nib}b) 
contradict the ``rule'' of only three NIB-values
in the purely deterministic model. 
Furthermore, for $T_S=0.55$ s ($T_V/T_S\approx 3$), 
fewer beats with NIB=2 are present in the data, 
and new NIB-values appear largely corresponding to the sequence
$5,8,11\ldots= 3n-1$, where $n\geq 2$ is an integer\cite{ctrig}.

There are also significant discrepancies between the predicted
fraction of V-beats from Eq.~(\ref{eq_V_freq}) and the observed
values. Fig.~\ref{V_freq}a shows that, for $T_V/T_S\approx 3$, 
the model of Eq.~(\ref{fde}) predicts many more V-beats than are found.

\def\figuraII{
\begin{figure}[t!]
\vspace*{.0cm}
\centerline{
\hspace*{0.0cm} 
\vbox{  
      \hbox{\epsfxsize=8cm \epsfbox{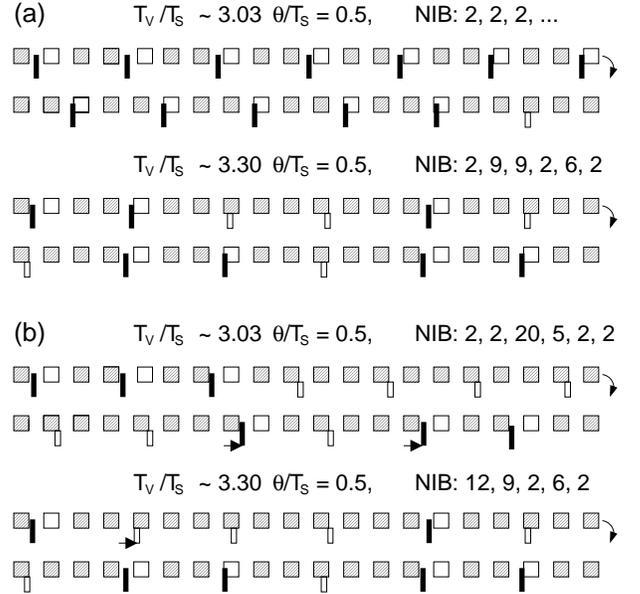}}
\vspace*{0.5cm}
      }
}
\caption{%
Generation of NIB-triplets
for $\theta/T_S=0.5$ and two different values of $T_V/T_S$.
The boxes are the S-beats including their refractory times. An empty
box is a blocked beat, and a dashed box an expressed beat. 
The expressed V-beats are symbolized by  black bars,
the blocked ones by empty bars.
The S-beat after each expressed abnormal beat is blocked
since $\theta/T_S\geq 0.5$.
(a) Deterministic model without coupling.
In the upper panel, the ratio of the periods $T_V/T_S \approx 3 $.
The phase of the V-beat with respect to the S-beats 
is therefore changing slowly, 
and the NIB is 2 for several periods
until the V-beat falls into the refractory time
of the S-beats and is blocked for several periods.
For $T_V/T_S\approx3.3$
 the NIB-triplet 2,6,9 is generated.
(b) Model with coupling and noise. 
Coupling shifts the V-beats into the refractory time
for  $T_V/T_S \approx 3$. 
For  $T_V/T_S\approx 3.3$ the coupling has minimal effects.
The addition of noise (arrows) changes the NIB-sequences
by moving some V-beats into or out of the refractory time.
For $T_V/T_S\approx 3$, 
this combines two consecutive appearances of NIB=2 into NIB=5 (8,11,$\ldots$).
For $T_V/T_S=3.3$,
combinations of NIB=2 with one of the other two values, 6 or 9,
generates NIB=9 or 12.}
\label{cartoon}
\end{figure}
}
\ifnum\tipo=1
  \figuraII
\fi
\ifnum\tipo=2
  \figuraII
\fi

We propose that the pattern of V-beats in the data
may be understood by assuming
that the timings of the V-beats are not strictly periodic as in
Eq.~(\ref{fde}), but that they are described by a stochastic
difference equation\cite{noise}
 \begin{equation}
\label{stochfde}
 \phi_{i+1} = \left(\phi_{i} + \frac{T_V}{T_S} + \frac{f(\phi_{i},T_S,T_V)}{T_S}
                                  + \frac{\eta}{T_S}\right)~{\rm mod}\,1,
\end{equation}
where $\eta$ is a Gaussian random variable distributed around 0,
and $f(\phi_i,T_S,T_V)$ gives the change of $T_V$ due to the coupling of the
S-beats to the V-beats\cite{moe,jalife,modulation}.
Each of the 2 or 3 S-beats that may appear between consecutive (expressed
or blocked) V-beats iteratively changes the position of the next V-beat
depending on the timing of the S-beats as shown in Fig.~\ref{V_freq}b. 
These changes add up to $f(\phi_{i},T_S,T_V)$\cite{formel}.

\def\figuraIII{
\ifnum\tipo=2
\end{multicols}
\fi
\begin{figure}
\centerline{
\vbox{  
    \hbox{
         \epsfxsize=5.6cm  \epsfbox{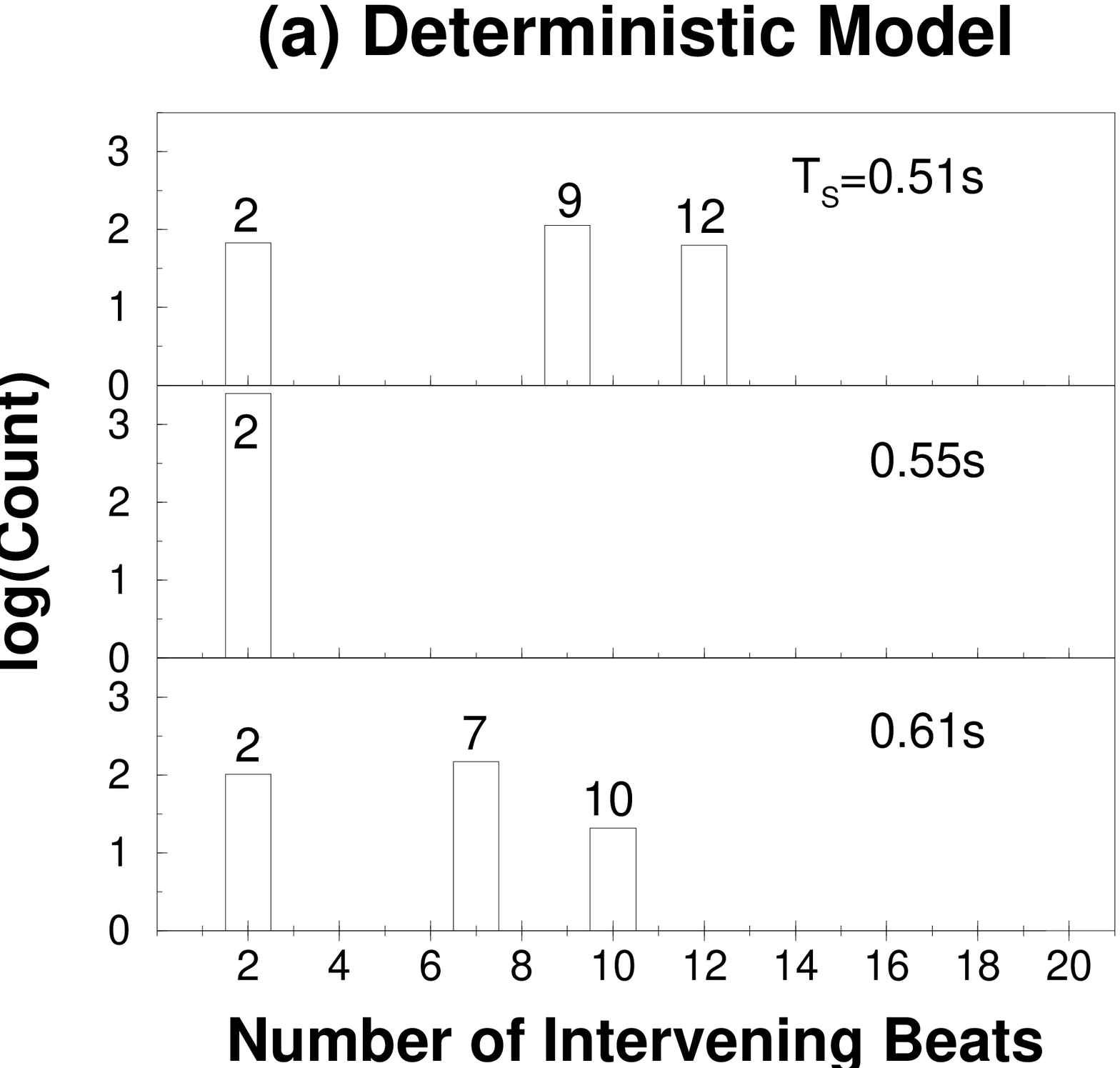}
         \hspace*{0.0cm}
         \epsfxsize=5.5cm  \epsfbox{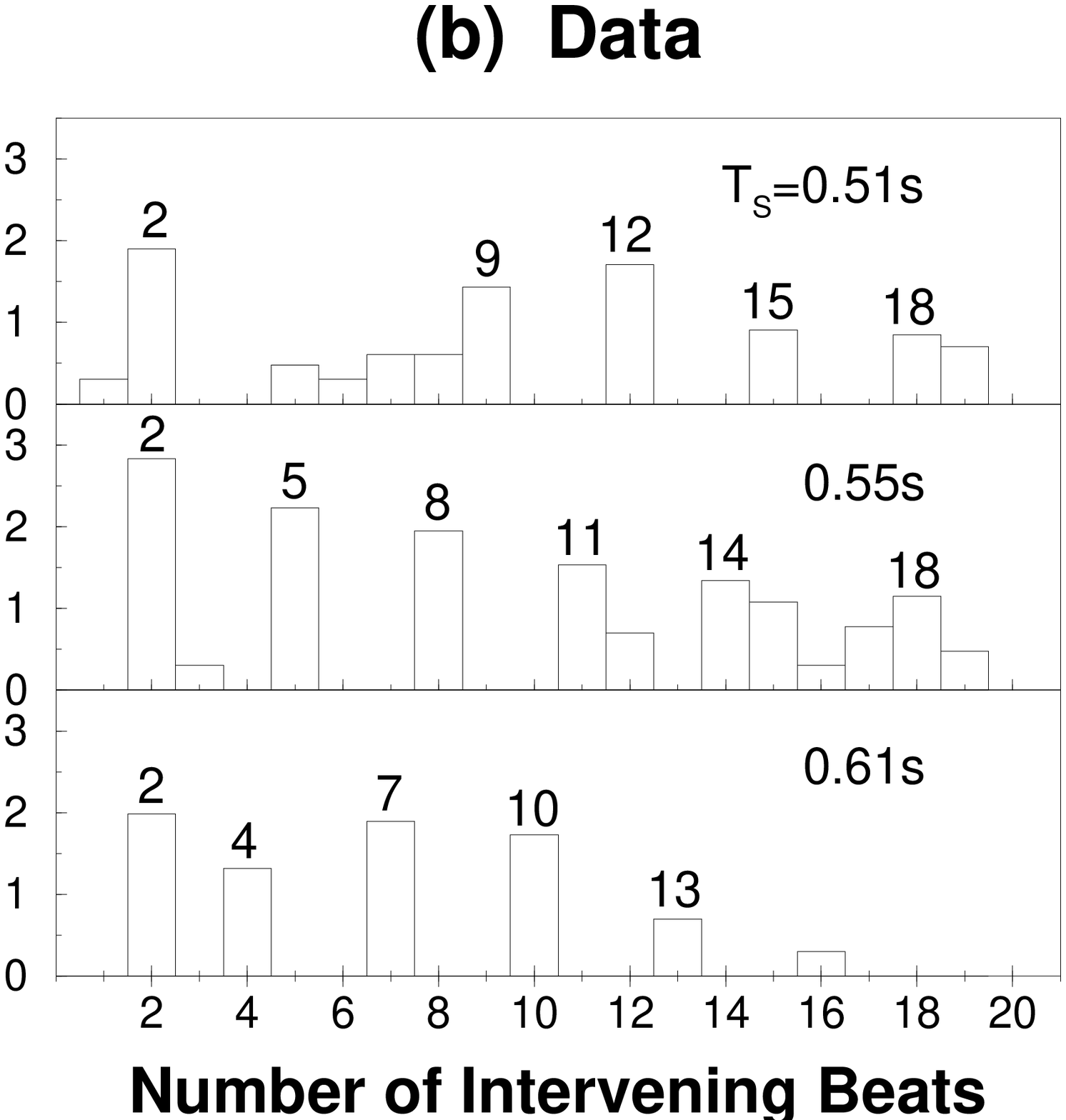}
         \hspace*{0.0cm}
         \epsfxsize=5.7cm  \epsfbox{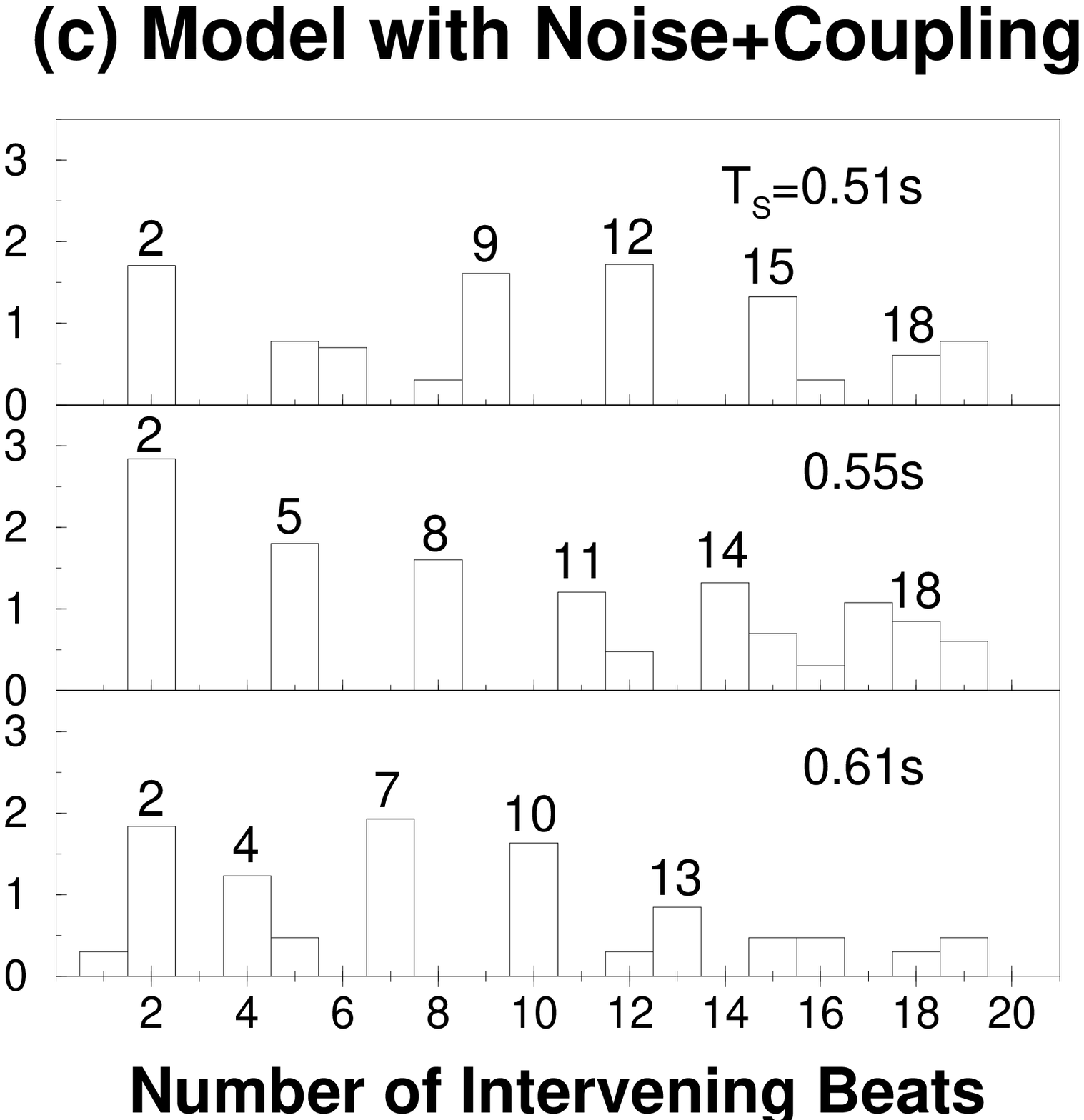}
          } 
       }
 }
\vspace*{0.5cm}
\caption{
 Histograms of the number of intervening beats (NIB) on a
 logarithmic scale for $T_S=0.51$~s, $0.55$~s, and $0.61$~s, 
 from top to bottom. For $T_V=$1.67~s, the corresponding ratios 
 of the periods are $T_V/T_S=$ 3.27, 3.03, and 2.73. 
 In the middle panel the count for NIB=2 is 2465 in (a), 
 677 in (b), and 686 in (c).
(a)  Simulation of the deterministic model with constant $T_S$, 
 $T_V=1.67$~s.
 The average S-beat interval $T_S$ varies from 0.47 s to 0.64 s
 based on the dataset we study.
 For each histogram, the number of S-beats corresponds to
 the number of S-beats in the data with the same average $T_S$.
(b) NIB-histogram for the data.
 The value for $T_S$  at each V-beat is the average S-beat interval in 
 a window of 20 beats centered at the V-beat.
 All V-beats in this window are replaced 
 by the corresponding blocked S-beats which are
 placed in the middle of the two neighboring S-beats.
 The histogram of the NIB-values is computed from all sequences of the
 record that have the same mean $T_S$.
 For $T_S=0.55$ s ($T_V/T_S$=3.03), 
 we find NIB-values $2,5,8,11,\ldots, 3n-1$
 that do not appear in the deterministic model.
(c) Simulation  with coupling (Fig.~\protect\ref{V_freq}b) 
 and noise. 
 Instead of a constant $T_S$, we use the S-beat intervals 
 of the data,  and we add to $T_V$ Gaussian white noise with
 a standard deviation $\sigma=0.07$~s.
 Note that the intrinsic period ${T_V}_0=1.76$ s is larger than the apparent 
 period $T_V\approx 1.67$s because of the coupling.}
\label{nib}
\end{figure}
\ifnum\tipo=2
\begin{multicols}{2}
\fi
}
\ifnum\tipo=1
  \figuraIII
\fi
\ifnum\tipo=2
  \figuraIII
\fi

In order to reproduce Fig.~\ref{V_freq}a we have chosen a
coupling\cite{curve} that leads, 
for $T_V/T_S \approx 3$, to a fixed point in Eq.~(\ref{stochfde}), 
such that the V-beats always fall in the refractory
period of the S-beat and thus are always blocked.
However, the noise term leads to a dispersion of the phases of the
V-oscillator such that some of the V-beats are
expressed. Thus, both the coupling and the stochastic term interact
to generate the dynamics.  The effects of the coupling and the noise
for $T_V/T_S \neq 3$ are represented schematically in
Fig.~\ref{cartoon}b. 
Fig.~\ref{nib}c shows how the
distributions of the NIB-values change when both coupling and noise
are included in the simulation.
The model now reproduces quantitatively the data. 
For $T_S \approx 0.55$~s ($T_V/T_S\approx 3$) the number of the
V-beats is reduced by the coupling, and the NIB-sequence $2,5,8,
\ldots$ is generated by the noise.
As a consequence of the random noise moving
the V-beats randomly in and out of the refractory time, the blocking
mechanism gives rise to a discrete Poisson process leading to an
approximately exponential fall-off of the peak heights of the
occurrences of the NIB values 2,5,8\cite{glassb,Longtin}.  
Finally, the model gives an accurate estimate of the fraction of
V-beats as a function of $T_S$ shown in Fig.~\ref{V_freq}a.

In order to estimate the magnitude of the noise, we calculate the
cross-correlation between the numbers of occurrences of each NIB value
in the model and the clinical data\cite{correlation}.  
Fig.~\ref{correl_fig} shows the correlation as a function of the
standard deviation $\sigma$ of the noise. The correlation function
has a maximum at $\sigma \approx 0.07$~s.  This value of $\sigma$ also best
reproduces the broadness of the peaks in the distribution of the
interventricular time intervals (Fig.~\ref{data}).  Further, the model
reproduces the asymmetrical form of 
\def\figuraIV{
\begin{figure}
\centerline{ 
\vbox{  
    \hbox{
        \epsfxsize=8cm \epsfbox{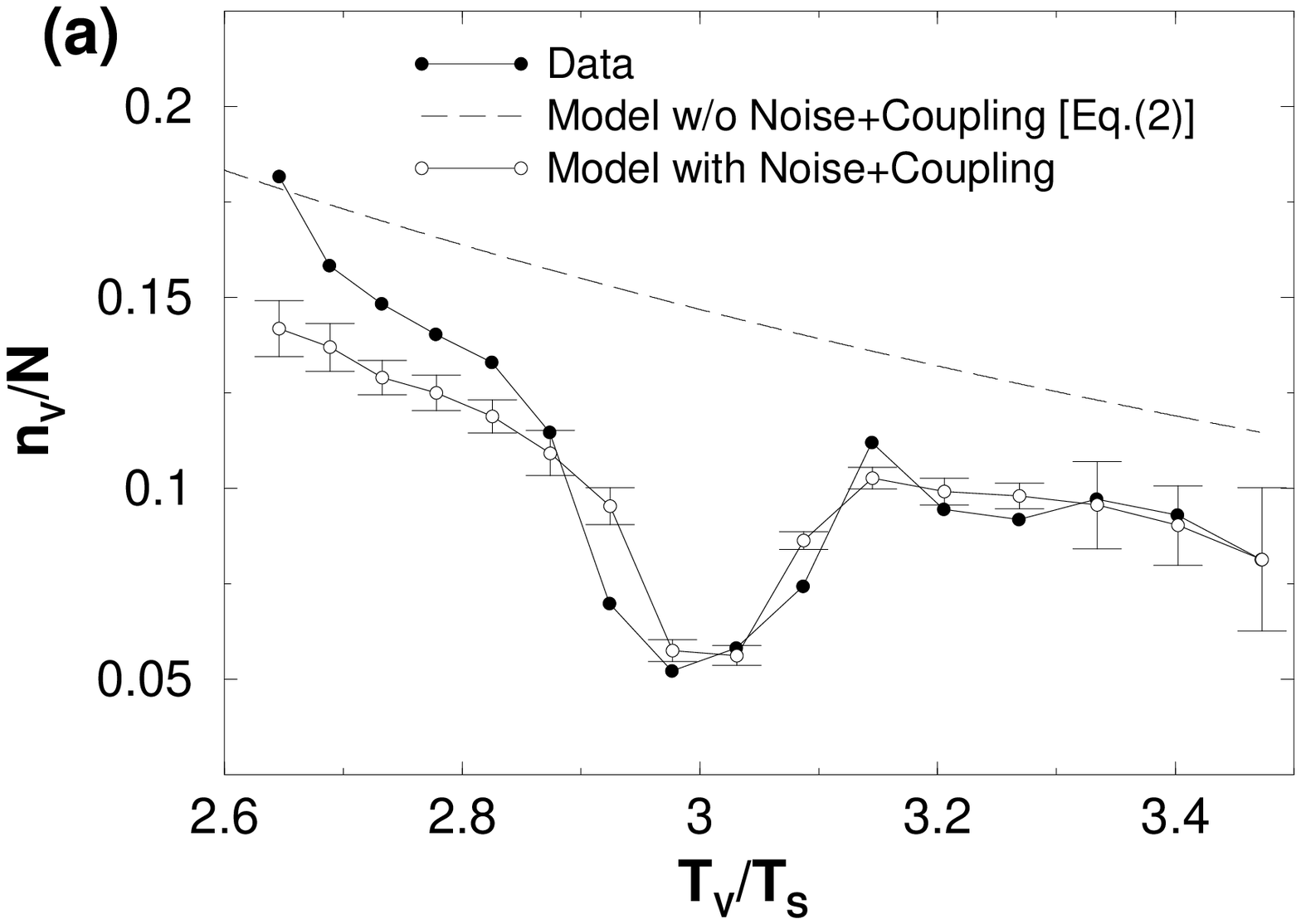}
         } 
}}	
\vspace*{-.2cm}
\centerline{ 
\vbox{  
    \hbox{
        \epsfxsize=7.5cm \epsfbox{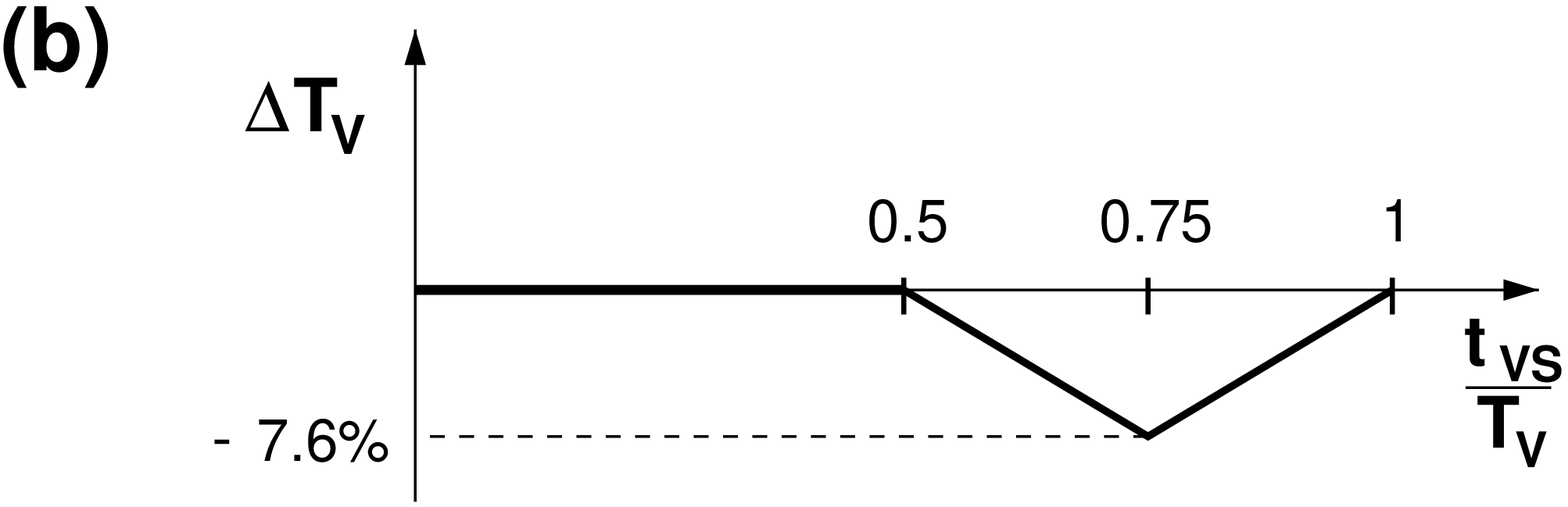}
         } 
}}	
\vspace*{0.1cm}
\caption{(a) The fraction $n_V/N$ plotted against the ratio $T_V/T_S$.
 The theoretical curve given by Eq.~(\protect\ref{eq_V_freq}), 
 with $\theta$ as in Fig.~\protect\ref{nib}a,
 is reproduced by the model without coupling, 
 and remains unchanged when noise is added.
 The data deviate from this curve at the ratio $T_V/T_S=3$,
 where we find less than half of the predicted V-beats.
 The model with coupling and noise reproduces this behavior.
 (b) The coupling between the two oscillators. 
 The change $\Delta T_V$ of the period of the V-oscillator
 as a function of the ratio of 
 the time $t_{\rm VS}$ between the last V-beat and an S-beat, and $T_V$.
 The coupling shortens the intrinsic ${T_V}_0$ to the apparent $T_V$.} 
\label{V_freq}
\end{figure}}
\ifnum\tipo=1
  \figuraIV
\fi
\ifnum\tipo=2
  \figuraIV
\fi
%
the peaks in the histogram
(inset of Fig. \ref{data}).  In the
simulation, the coupling splits the peaks into two subpeaks giving
them an asymmetrical appearance (not shown).

In this work we analyzed the patterns of abnormal heartbeats in 
a 12 hour record from a single patient and 
proposed that the dynamics results from a 
combination of deterministic and stochastic mechanisms. 
Quantitative comparison between predictions of the model
and the clinical data shows best agreement for an optimal level 
of noise in the model. 

Our approach is in contrast to standard approaches\cite{echt} 
in which crude measures, such as the average numbers 
of abnormal heart beats per unit time, 
are used for clinical assessment.  
We believe that distinctive dynamics are associated 
with different mechanisms, 
and hence different therapeutic strategies. 
Thus, the detailed program of analysis 
applied to one clinical record in this Letter 
is essential to better classify cardiac arrhythmias
based on the underlying mechanisms and resulting dynamics.

\def\figuraV{
\begin{figure}[t!]
\centerline{ 
\vbox{
    \hbox{
        \epsfxsize=8cm \epsfbox{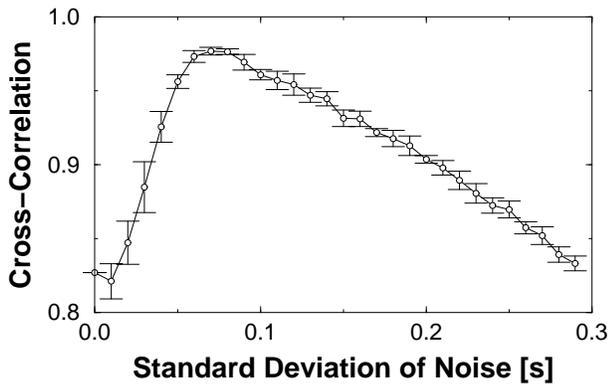}
         } 
      }
}
\caption{The cross-correlation between data and model
 as a function of noise. We cross-correlate the histograms in 
 Fig.~\protect\ref{nib}b,c for all values of $T_S$ 
 from 0.48-0.62 s in steps of 0.01s~\protect\cite{correlation}.}
\label{correl_fig}
\end{figure}}
\ifnum\tipo=1
  \figuraV
\fi
\ifnum\tipo=2
  \figuraV
\fi

We thank R.~Goldsmith for providing the data, 
and the German Academic Exchange Service (DAAD),
NIH/NCRR (P41RR13622), NASA, and the Mathers Charitable Foundation
for support.



\ifnum \tipo=3
  \newpage
  \figuraI
\fi  

\ifnum \tipo=3
 \newpage
 \figuraII
\fi

\ifnum \tipo=3
 \newpage        
 \figuraIII
\fi

\ifnum \tipo=3
 \newpage
 \figuraIV
\fi

\ifnum \tipo=3
 \newpage
 \figuraV
\fi

\ifnum \tipo=2
\end{multicols}
\fi


\begin{references}
\bibitem{echt} D.S.~Echt {\it et al.}, 
 New Engl.~J.~Med.\ {\bf 324}, 781 (1991).
\bibitem{moe} 
 G.K. Moe {\it et al.}, Circulation {\bf 56}, 968 (1977).
\bibitem{courte} 
 M.~Courtemanche {\it et al.}, Am.\ J.\ Physiol.~{\bf 257}, H693 (1989); 
 Physica D {\bf 40}, 299 (1989).
\bibitem{jalife} 
 J. Jalife {\it et al.}, Pace {\bf 5}, 911 (1982);
 N.~Ikeda {\it et al.}, J.~Theor.~Biol.~{\bf 103}, 439 (1983). 
\bibitem{glass} 
 L. Glass {\it et al.}, Am. J. Physiol. {\bf 251}, H841 (1986). 
\bibitem{glassb}L. Glass {\it et al.}, Proc. R. Soc. Lond. A  {\bf 413}, 9 (1987).
\bibitem{depaola}
 R. De Paola {\it et al.}, Am. J. Physiol. {\bf 265}, H1603 (1993);
 H.-X. Wang {\it et al.}, Phys.\ Rev.\ Lett.\ {\bf 70}, 3671 (1993);
 H.-X. Wang {\it et al.}, {\it ibid.} {\bf 71}, 3039 (1993).
\bibitem{liebovitch}
 L.S. Liebovitch {\it et al.}, Phy.~Rev.~E {\bf 59}, 3312 (1999).
\bibitem{babloyantz}
 A.~Babloyantz and P.~Maurer, Phys.\ Lett.\ A {\bf 221}, 43 (1996).
\bibitem{refract} The refractory times after the S- and the V-beats might, 
 in fact, be different. The refractory time after the S-beats is
 approximated as the minimal time interval between an S-beat and the
 following V-beat. Analysis (not given here) shows that the refractory
 time shows a weak dependence on $T_S$. In the following we assume
 $\theta=0.29T_S+0.15$ s.  The refractory time after the V-beats
 cannot (and need not) be estimated in the same way since in this
 case, the following S-beats are always blocked.
\bibitem{rules}
 For the purely deterministic model, the NIB-sequences 
 can be shown to follow three ``rules'':~\protect\cite{courte,glass}:
 (i) there are at most three different values for the NIB;
 (ii) the sum of the two smaller ones is the largest NIB-value minus one,
 implying that
 (iii) one of the NIB-values must be odd.
 The $\theta/T_S,\ T_V/T_S$-parameter space can be completely
 partitioned into regions of these triplets\protect\cite{glass}. 
 The problem is related to the gaps and steps problem
 in number theory:
 G.A.~Hedlund, Am.~J.~Math.~{\bf 66}, 605 (1944);
 N.B.~Slater, Proc.\ Camb.\ Phil.\ Soc.\ {\bf 63}, 1115 (1967). 
\bibitem{ctrig}Such a rhythm is called concealed trigeminy:
 L.\ Schamroth, H.J.L.\ Marriott, Am.\ J.\ Cardiol.\ {\bf 7}, 799
 (1961).
\bibitem{noise}
  A.~Lasota and M.C.~Mackey, 
 {\it Chaos, Fractals and Noise: Stochastic Aspects of Dynamics}, 
 Springer Verlag, New York (1994). 
\bibitem{modulation}In the clinical literature such a coupling effect
 is referred to as modulated parasystole.
\bibitem{formel}
 We here give an expression for $f(\phi_{i},T_S,T_V)$.
 Each S-beat between two consecutive V-beats 
 changes the time interval between the V-beats.
 The time interval $t^n_{\rm VS}$ between a V-beat with the phase
 $\phi_i$ and the $n$th S-beat is $(n-\phi_i)\,T_S$, 
  $n=1,\ldots , n_{\rm max}$, 
 where the S-beat with $n=1$ is blocked if the V-beat was expressed.
 The period ${T_V}_{n}$ including the effects of  all S-beats 
 up to the $n$th is found by iterating:
 \[
 {T_V}_{n}={T_V}_{n-1}
         +\Delta T_V \!\left(\frac{(n-\phi_i)\, T_S}
                                  {{T_V}_{n-1}}
                       \right),
 \]
 where $\Delta T_V(x)$ is given by Fig.~\protect\ref{V_freq}b
 and ${T_V}_0=1.76$~s, the intrinsic period of the V-beats.
 In Eq.~(\protect\ref{stochfde}), 
 we then have $f(\phi_i,T_S,T_V)={T_V}_{n_{\rm max}}-T_V$. 
 The number $n_{\rm max}$ of S-beats between 
 consecutive  V-beats, expressed or blocked,
 is the smallest integer number such that 
 $(n_{\rm max}-\phi_i)T_S \leq {T_V}_{n_{\rm max}}$.
 In our particular case, $n_{\rm max}$ may be 3 or 4.
\bibitem{curve}   Some other coupling curves are possible 
  and yield similar agreement with the data.
  The coupling curves with best fit to our data
  all had a flat or positive first part ($T_V/T_S\leq 0.5$) 
  and a negative second part ($T_V/T_S\geq 0.5$).
  This form is similar to physiologically 
  motivated phase resetting curves \protect\cite{moe,jalife}.
\bibitem{Longtin}
 A.\ Longtin, Chaos {\bf 5}, 209 (1995).
\bibitem{correlation}
 The cross-correlation is calculated with: 
 $\big[\sum_{ij}\!(x_{ij}\!-\!\overline{x})\\(y_{ij}\!-\overline{y})\big]/
       \big[\sum_{ij}\! (x_{ij}\!-\overline{x})^2\, 
         \sum_{ij}\! (y_{ij}\!-\overline{y})^2\big]^{1/2}\!$,
 where  $x_{ij}$ and $y_{ij}$ are the components of the matrices
 in Fig.~\protect\ref{nib}b and c. 
\end{references}
\end{document}